\documentclass[conference]{IEEEtran}
\IEEEoverridecommandlockouts
\usepackage{cite}
\usepackage{amsmath,amssymb,amsfonts}
\usepackage{mathtools}
\usepackage{algorithm}
\usepackage{algpseudocode}

\usepackage{graphicx}
\usepackage{textcomp}
\usepackage{xcolor}
\def\BibTeX{{\rm B\kern-.05em{\sc i\kern-.025em b}\kern-.08em
    T\kern-.1667em\lower.7ex\hbox{E}\kern-.125emX}}

\makeatletter
\def\endthebibliography{%
	\def\@noitemerr{\@latex@warning{Empty `thebibliography' environment}}%
	\endlist
}
\makeatother

\usepackage{amsthm}
\newtheorem{definition}{Definition}

\makeatletter
\newcommand{\removelatexerror}{\let\@latex@error\@gobble}
\makeatother

\usepackage{tikz}
\usepackage{pgfplots} 
\pgfplotsset{compat=newest}
\usetikzlibrary{spy,arrows,positioning,shapes,shapes.geometric,external,shadows,babel,calc}
 
\ifCLASSOPTIONcompsoc
\usepackage[caption=false, font=normalsize, labelfont=sf, textfont=sf]{subfig}
\else
\usepackage[caption=false, font=footnotesize]{subfig}
\fi

\newcommand\copyrighttext{%
	\footnotesize \copyright\,2020 IEEE. Personal use of this material is permitted. Permission from IEEE must be obtained for all other uses, in any current or future media, including reprinting/republishing this material for advertising or promotional purposes, creating new collective works, for resale or redistribution to servers or lists, or reuse of any copyrighted component of this work in other works.}%
\newcommand\copyrightnotice{%
	\begin{tikzpicture}[remember picture,overlay]%
	\node[anchor=south,yshift=10pt] at (current page.south) {\fbox{\parbox{\dimexpr\textwidth-2cm}{\copyrighttext}}};%
	\end{tikzpicture}%
	\vspace{-10pt}%
}

\begin{document}

\title{Kalman Filter Meets Subjective Logic:\\ A Self-Assessing Kalman Filter Using Subjective Logic
\thanks{This research is accomplished within the project SecForCARs (grant number 16KIS0795). We acknowledge the financial support for the project by the Federal Ministry of Education and Research of Germany (BMBF).}
}

\author{\IEEEauthorblockN{Thomas Griebel, Johannes M\"uller, Michael Buchholz, and Klaus Dietmayer}
\IEEEauthorblockA{\textit{Institute of Measurement, Control and Microtechnology} \\
\textit{Ulm University}\\
89081 Ulm, Germany \\
\{thomas.griebel, johannes-christian.mueller, michael.buchholz, klaus.dietmayer\}@uni-ulm.de}
}

\maketitle
\copyrightnotice

\begin{abstract}
	
Self-assessment is a key to safety and robustness in automated driving.
In order to design safer and more robust automated driving functions, the goal is to self-assess the performance of each module in a whole automated driving system.  
One crucial component in automated driving systems is the tracking of surrounding objects, where the Kalman filter is the most fundamental tracking algorithm.
For Kalman filters, some classical online consistency measures exist for self-assessment, which are based on classical probability theory.
However, these classical approaches lack the ability to measure the explicit statistical uncertainty within the self-assessment, which is an important quality measure, particularly, if only a small number of samples is available for the self-assessment.
In this work, we propose a novel online self-assessment method using subjective logic, which is a modern extension of probabilistic logic that explicitly models the statistical uncertainty. 
Thus, by embedding classical Kalman filtering into subjective logic, our method additionally features an explicit measure for statistical uncertainty in the self-assessment. 

\end{abstract}


\section{Introduction}

Being already widely used in the field of avionics and navigation~\cite{pullen2011}, monitoring and assuring systems' functional performance have recently gained more and more importance for automated vehicles and is generally termed \textit{safety of the intended functionality} (SOTIF) in the automotive context. 
Thus, self-assessment of the individual modules plays an important role to reach SOTIF; see, e.g.,~\cite{mueller2019}. One crucial module in the perception of automated vehicles is the tracking of objects in its surrounding environment. 
For this task, the Kalman filter~\cite{kalman1960new} is the most fundamental algorithm.

Classical approaches use the well-known \textit{normalized innovation squared} (NIS)~\cite{bar1988tracking} for online self-assessment of Kalman filtering. 
The NIS monitors whether the Kalman filter's noise assumptions are consistent with the incoming measurements.
In~\cite{gibbs2013new}, Gibbs presents three tests to examine inconsistencies in Kalman filtering. The tests are designed to detect measurement outliers and model inconsistencies.
Similar self-assessment quality measures have recently been used to adapt Kalman filter parameters depending on changing environments~\cite{gelen2018new,chen2018weak,chen2019kalman}.
However, none of these works has taken into account the statistical uncertainty of the quality measure used for self-assessment.
More precisely, the statistical uncertainty explicitly expresses the confidence of the quality measure itself. This type of uncertainty is typically called second-order probability.
In fact, the statistical uncertainty can play an important role in self-assessment, particularly, if the number of samples is limited such that the quality measures may have limited statistical meaning. Then, using the statistical uncertainty, we are able to directly take into account how long the filter has already been consistent regarding the incoming measurements. This additional information can be further used to improve overall performance of the filter.

\begin{figure}[!t]
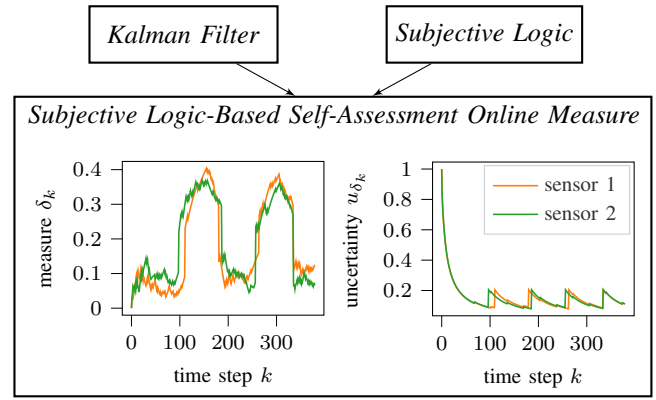

	\centering
		\begin{tikzpicture}[>=latex']
		\tikzset{block/.style= {draw, rectangle, align=center,minimum width=2.5cm,minimum height=0.75cm},
		}
		\node (start) at (0,0) {\textit{Subjective Logic-Based Self-Assessment Online Measure}};
		\node [block, thick, above of=start, xshift=2cm, yshift=0.1cm] (start01){\textit{Subjective Logic}};
		\node [block, thick, above of=start, xshift=-2cm, yshift=0.1cm] (start02){\textit{Kalman Filter}};
		\node[inner sep, below of=start, xshift=-2.1cm, yshift=-1.1cm] (dc) {\centering\input{resources/figures/exp03/dc_gauss_final_title.tex}};
		\node[inner sep, below of=start, xshift=2cm, yshift=-1.1cm] (uncertainty) {\centering\input{resources/figures/exp03/uncertainty_gauss_final_title.tex}};
		
		\path[draw, ->]
		(start01) edge (start)
		(start02) edge (start)
		;
		\draw[thick] ($(dc.north west)+(-0.05,0.64)$) rectangle ($(uncertainty.south east)+(0.1,+0.125)$);
		
		\end{tikzpicture}
	\caption{Concept of our proposed subjective logic-based self-assessment method in Kalman filtering. \label{fig:concept_sl-based_self_assessment}}
\end{figure}

In this work, we present a novel approach to obtain a self-assessment measure in Kalman filtering using subjective logic; see Fig.~\ref{fig:concept_sl-based_self_assessment}.
Subjective logic is a mathematical theory that explicitly models statistical uncertainty~\cite{Joesang2016} similar to the Dempster-Shafer theory~\cite{shafer1976mathematical, dempster1967upper}. Thus, our approach features a reliability measure that explicitly includes statistical uncertainty. This additional measure can be particularly beneficial if the number of samples is strictly limited, e.g., due to a fast-changing environment as we often observe in automated driving. Our proposed self-assessment method is able to online estimate the Kalman filter's performance and is presented as closed-form implementation in the theory of subjective logic.

Our contribution is two-fold: from a theoretical perspective, this work creates a never-before-seen link between subjective logic theory and Kalman filtering.
From a practical perspective, we introduce a new online quality measure for self-assessment of Kalman filtering that additionally features a measure for the statistical uncertainty. 

The remainder of this work is structured as follows. 
Section~\ref{section:RelatedWork} describes similar works in the related field. In Section~\ref{section:basics}, the fundamentals of subjective logic and Kalman filtering are summarized. Section~\ref{section:self_assessing_method} presents our proposed method to obtain a self-assessment online measure for Kalman filtering using subjective logic. The simulation results of our proposed method are discussed in Section~\ref{section:experiments}. Finally, Section~\ref{section:conclusion} concludes our work.

\section{Related Work} \label{section:RelatedWork}

The classical quality measure in Kalman filtering is the NIS~\cite{bar1988tracking}. Based upon the NIS and the \textit{normalized estimation error squared} (NEES)~\cite{bar1988tracking}, which needs, in contrast to the NIS, ground truth data, further consistency measures have been introduced in recent years.
In~\cite{gibbs2013new}, Gibbs presents three tests to examine inconsistencies in Kalman filtering. The smoother residual test and smother state test are derived, which are both based on a modified Bryson-Frazier smoother and are designed to detect measurement outliers and model inconsistencies, respectively.
In addition, a filter residual test is introduced, which is also designed to detect measurement outliers.
In~\cite{piche2016online}, three equivalent derivations of the NIS and the resulting evaluation alternatives are presented. Firstly, the NIS is derived as a Bayesian p-test for the prior predictive distribution. Secondly, a derivation as a nested-model parameter significance test is given. Thirdly, a filter residual approach is described. 
In~\cite{gamse2014statistical}, a detailed evaluation of Kalman filtering is presented including indicators of, e.g., inner confidence, the determinant of the state transition matrix, properties of covariance matrices, and the Kalman gain.

Furthermore, Kalman filter tuning and adaptive Kalman filtering, which is often based on consistency measures, have gained some research attention in recent years. In~\cite{scalzo2009adaptive}, adaptive filtering for single target tracking is proposed, which selects appropriate filter algorithms depending on the NIS. 
Gelen et al.~\cite{gelen2018new} develop three metrics to tune the Kalman filter in terms of process noise and measurement noise parameters.
In~\cite{chen2018weak}, a method for auto-tuning Kalman filters with a Bayesian optimization strategy based on the NIS and NEES is designed. This method, however, needs ground truth data in order to use the NEES.
Recently, Chen et al.~\cite{chen2019kalman} present how Bayesian optimization can resolve some issues in parameter tuning of Kalman filtering without having ground truth data.

In the context of temporal filtering and subjective logic, a subjective logic-based identification of Markov chains has been developed in~\cite{muller2019subjective}. The presented identification method generates, in addition to classical approaches, an explicit reliability measure in terms of statistical uncertainty of the identification result itself. 
Only slightly related is the approach of {\v{S}}kori{\'c} et al.~\cite{vskoric2016flow}. They present evidence-based subjective logic as a combination of flow-based reputation systems with the uncertainty concept of subjective logic in order to determine indirect computational trust through a trust network. In fact, flow-based reputation systems have their mathematical foundation also in Markov chains.

However, to the best of our knowledge, neither the combination of subjective logic and Kalman filtering, nor the introduction of a self-assessment metric for Kalman filtering that explicitly includes a measure for the statistical uncertainty have been addressed in literature so far.

\section{Fundamentals} \label{section:basics}

This section summaries the mathematical foundation of subjective logic including some commonly used subjective logic operators, which are also required for our proposed method. In addition, we briefly summarize the Kalman filter and outline the consistency examination of Kalman filtering.

\subsection{Subjective Logic}
The mathematical description of subjective logic, which is summarized in the following, is mainly based on \cite{Joesang2016}. One key structure in subjective logic is the opinion representation. A multinomial opinion expresses information of a discrete random variable $X$ in terms of belief, uncertainty, and base rate for every event $x$ of the sample space $\mathbb{X}$.
\begin{definition}[Multinomial Opinion]
	Let $X \in \mathbb{X}$ be a random variable of the finite domain $\mathbb{X}$ with cardinality $W = |\mathbb{X}| \geq 2$. A multinomial opinion is an ordered triple $\omega_X = (\boldsymbol{b}_X,u_X,\boldsymbol{a}_X)$ with
	\begin{subequations}
		\begin{align}
		\boldsymbol{a}_X(x) : \mathbb{X} \mapsto [0,1], \qquad 1 &= \sum\limits_{x \in \mathbb{X}} \boldsymbol{a}_X (x) \, ,\\
		\boldsymbol{b}_X(x) : \mathbb{X} \mapsto [0,1], \qquad 1 &= u_X + \sum\limits_{x \in \mathbb{X}} \boldsymbol{b}_X (x)\, .
		\end{align}
	\end{subequations}
	Here, $\boldsymbol{b}_X$ is the belief mass distribution over $\mathbb{X}$, $u_X \in [0,1]$ is the uncertainty mass representing the lack of evidence, and $\boldsymbol{a}_X$ is the base rate distribution over $\mathbb{X}$ representing the prior probability.
	Moreover, the projected probability distribution
	\begin{equation}
	\boldsymbol{P}_X(x) = \boldsymbol{b}_X(x) + \boldsymbol{a}_X(x) u_X, \quad \forall x \in \mathbb{X},
	\end{equation}  
	of a multinomial opinion projects the opinion to a classical probability distribution and, thus, represents the expected outcome of an opinion in probability space.
\end{definition}
To combine opinions from various sources about the same domain of interest, multiple fusion operators exist for merging these opinions. Generally speaking, this can be interpreted as a set of sources that come together in order to find a joint conclusion about a certain task using some fusion operator. For certain tasks, particular fusion operators are more suitable than others. Here, we present the \textit{aleatory cumulative belief fusion} (A-CBF), which is an appropriate fusion operator for our method in Section~\ref{section:self_assessing_method}. Further fusion operators can be found in~\cite{Joesang2016}.
\begin{definition}[Aleatory Cumulative Belief Fusion]
	Let $\omega_X^A$ and $\omega_X^B$ be multinomial opinions of source A and B over the same variable $X$ on domain $\mathbb{X}$. Let $\omega_X^{A \diamond B}$ be the fused opinion such that
	\begin{equation} \label{eqn:a-cbf}
	\omega_X^{A \diamond B} = \left \{  \begin{array}{ll}
	\boldsymbol{b}_X^{A \diamond B}(x) &= \frac{ \boldsymbol{b}_X^{A}(x) u_X^{B} + \boldsymbol{b}_X^{B}(x) u_X^{A} }{ u_X^{A} + u_X^{B} - u_X^{A} u_X^{B} } \\
	\\[-6pt]
	u_X^{A \diamond B} &= \frac{ u_X^{A} u_X^{B} }{ u_X^{A} + u_X^{B} - u_X^{A} u_X^{B} }\\
	\\[-6pt]
	\boldsymbol{a}_X^{A \diamond B}(x) &= \frac{ \boldsymbol{a}_X^{A}(x) u_X^{B} + \boldsymbol{a}_X^{B}(x) u_X^{A} }{ u_X^{A} + u_X^{B} - 2 u_X^{A} u_X^{B} } 
	\\[6pt]
	&\quad - \frac{ (\boldsymbol{a}_X^{A} (x) + \boldsymbol{a}_X^{B}(x)) u_X^{A} u_X^{B} }{ u_X^{A} + u_X^{B} - 2 u_X^{A} u_X^{B} } 	 	
	\end{array}\right .
	\end{equation}
	for $u_X^A \neq 0$ $\lor$ $u_X^B \neq 0$ and $u_X^A \neq 1$ $\lor$ $u_X^B \neq 1$, then the operator $\oplus$ in $\omega_X^{A \diamond B} = \omega_X^A \oplus \omega_X^B$ is called aleatory cumulative belief fusion. For special cases as $u_X^A = u_X^B = 0$ or $u_X^A = u_X^B = 1$, we refer to \cite{Joesang2016}.
\end{definition}
The opposite of fusion in subjective logic is called unfusion. The objective of an unfusion operator is to remove the input of a specific opinion from an already fused opinion. In fact, the unfusion operator of the A-CBF is called cumulative unfusion~\cite{Joesang2016}.
\begin{definition}[Cumulative Unfusion]
	Let $\omega_X^C = \omega_X^{A \diamond B}$ be the cumulative fused opinion as in~\eqref{eqn:a-cbf} of $\omega_X^B$ and an unknown opinion $\omega_X^A$ over the variable $X \in \mathbb{X}$ with the same base rate of opinion $B$ and $C$, namely $\boldsymbol{a}_X$. Let $\omega_X^A = \omega_X^{C \bar{\diamond} B}$ be the unfused opinion such that
	\begin{equation}
	\omega_X^{C \bar{\diamond} B} = \left \{  \begin{array}{ll}
	\boldsymbol{b}_X^{C \bar{\diamond} B}(x) &= \frac{ \boldsymbol{b}_X^{C}(x) u_X^{B} - \boldsymbol{b}_X^{B}(x) u_X^{C} }{ u_X^{B} - u_X^{C} + u_X^{B} u_X^{C} } \\
	\\[-6pt]
	u_X^{C \bar{\diamond} B} &= \frac{ u_X^{B} u_X^{C} }{ u_X^{B} - u_X^{C} + u_X^{B} u_X^{C} }\\
	\\[-6pt]
	\boldsymbol{a}_X^{C \bar{\diamond} B}(x) &= \boldsymbol{a}_X(x)	 	
	\end{array}\right .
	\end{equation}
	for $u_X^B \neq 0$ $\lor$ $u_X^C \neq 0$, then the operator $\ominus$ in $\omega_X^{C \bar{\diamond} B} = \omega_X^C \ominus \omega_X^B$ is called cumulative belief unfusion. For the special case $u_X^B = u_X^C= 0$, we refer to~\cite{Joesang2016}.
\end{definition}
To obtain trust or belief from transitive trust paths, trust discounting is often used; for further details, please refer to~\cite{Joesang2016}. We define and use trust discounting in a different way for our purpose.
\begin{definition}[Trust Discounting] 
	Let $\omega_X^A$ be source A's opinion over $X$ on domain $\mathbb{X}$ and $p_d \in [0, 1]$ be the discount probability.
	Then, with $\emph{TD}\left(\omega_X^{A}, p_d\right)$ denoting trust discounting of opinion $\omega_X^{A}$ with respect to $p_d$, let $\omega_X^{A_{p_d}} =\emph{TD}\left(\omega_X^{A}, p_d\right)$ be the trust discounted opinion such that
	\begin{equation}
	\label{eqn:TrustDiscounting}
	\omega_X^{A_{p_d}} = \left \{ \begin{array}{ll}
	\boldsymbol{b}_X^{A_{p_d}}(x) &= p_{d} \; \boldsymbol{b}_X^{A}(x) \\
	u_X^{A_{p_d}} &= 1 - p_{d} \sum \limits_{x \in \mathbb{X}} \!  \boldsymbol{b}_X^{A}(x) \\
	\boldsymbol{a}_X^{A_{p_d}}(x) &= \boldsymbol{a}_X^{A}(x) 
	\end{array}\right . \, .
	\end{equation}
\end{definition}
Roughly speaking, trust discounting models that a certain amount of information will be lost while transferring this information via multiple sources. We use this subjective logic operator in our proposed method in the context of estimating time-varying parameters. More precisely, we use trust discounting to account for information degradation over time due to possible parameter changes.

Apart from fusion operators, a comparison operator called \textit{degree of conflict} (DC) is defined in order to measure the difference between two opinions about the same variable $X$. 
\begin{definition}[Degree of Conflict]
	Let $\omega_X^A$ and $\omega_X^B$ be multinomial opinions of source A and B over the same variable $X$ on domain $\mathbb{X}$. Then, $\emph{DC} \left(\omega_X^A, \omega_X^B\right)$ denotes the degree of conflict between the two opinions $\omega_X^A$ and $\omega_X^B$. The $\emph{DC}$ is defined as
		\begin{equation}
		\label{eqn:DC}
		\emph{DC} \left(\omega_X^A, \omega_X^B\right) = \emph{PD} \left(\omega_X^A, \omega_X^B\right) \cdot \emph{CC} \left(\omega_X^A, \omega_X^B\right) ,
		\end{equation} 
	where ${\emph{PD} \left(\omega_X^A, \omega_X^B\right) = \frac{1}{2} \sum_{x \in \mathbb{X}} |\boldsymbol{P}_{X}^A (x) - \boldsymbol{P}_{X}^B (x)| \in \left[0, 1\right]}$ denotes the projected distance and $\emph{CC} \left(\omega_X^A, \omega_X^B\right) = \left( 1 - u_X^A\right) \left( 1 - u_X^B\right) \in \left[0, 1\right]$ the conjunctive certainty.
\end{definition}
Obviously, it holds that $\text{DC} \in \left[0, 1\right]$. For similar opinions, the $\text{DC}$ is expected to be small, i.e., nearly zero, and for highly conflicting opinions, the $\text{DC}$ is expected to be large, i.e., nearly the value of $\text{CC}$.

\subsection{Kalman Filter}

The Kalman filter~\cite{kalman1960new} is an estimation algorithm for unknown variables based on a series of uncertain measurements.
The key assumptions of Kalman filtering are that all signals and probability densities are Gaussian distributed and the process and measurement models are linear. If these assumptions are fulfilled, then the Kalman filter is a Bayes-optimal state estimator~\cite{ho1964bayesian} and facilitates a closed-form implementation of the Bayes filter for recursive state estimations.

The estimated state $\boldsymbol{x}_k \in \mathbb{R}^{n}$ of an object at time step $k \in \mathbb{N}$ in Kalman filtering is modeled by an $n$-dimensional multivariate Gaussian distribution with mean $\hat{\boldsymbol{x}}_k \in \mathbb{R}^{n}$ and covariance matrix $\boldsymbol{P}_k \in \mathbb{R}^{n \times n}$.
The motion and measurement models are given by
\begin{align}
\boldsymbol{x}_{k+1} &= \boldsymbol{F}_k \boldsymbol{x}_{k} + \boldsymbol{v}_{k} , \label{eqn:motion_model}\\
\boldsymbol{z}_{k} &= \boldsymbol{H}_k \boldsymbol{x}_{k} + \boldsymbol{w}_{k}
\end{align} 
with the process matrix $\boldsymbol{F}_k \in \mathbb{R}^{n \times n}$ and the measurement matrix $\boldsymbol{H}_k \in \mathbb{R}^{m \times n}$. The process noise $\boldsymbol{v}_{k}\in \mathbb{R}^{n}$ and measurement noise $\boldsymbol{w}_{k} \in \mathbb{R}^{m}$ are assumed to be uncorrelated and zero-mean Gaussian distributed. Then, the motion model in~\eqref{eqn:motion_model} yields to the predicted state of the object with the corresponding covariance matrix
\begin{align}
\hat{\boldsymbol{x}}_{k+1|k} &= \boldsymbol{F}_k \hat{\boldsymbol{x}}_{k} , \\
\boldsymbol{P}_{k+1|k} &= \boldsymbol{F}_k \boldsymbol{P}_k \boldsymbol{F}_k^T + \boldsymbol{Q}_{k} ,
\end{align}
where $\boldsymbol{Q}_{k} = \mathbb{E} \left[ \boldsymbol{v}_{k} \boldsymbol{v}_{k}^T \right] \in \mathbb{R}^{n \times n}$  is the covariance matrix of the process noise. The measurement prediction is stated by
\begin{align}
\hat{\boldsymbol{z}}_{k+1|k} &= \boldsymbol{H}_{k+1} \hat{\boldsymbol{x}}_{k+1|k} , \label{eqn:kf_meas_pred}\\
\boldsymbol{S}_{k+1} &= \boldsymbol{H}_{k+1} \boldsymbol{P}_{k+1|k} \boldsymbol{H}_{k+1}^T + \boldsymbol{R}_{k+1} , \label{eqn:kf_meas_pred_cov}
\end{align}
where $\boldsymbol{R}_{k+1} = \mathbb{E} \left[ \boldsymbol{w}_{k+1} \boldsymbol{w}_{k+1}^T \right] \in \mathbb{R}^{m \times m}$ is the covariance matrix of the predicted measurement. Thus, the measurement matrix $\boldsymbol{H}_{k+1}$ displays the transformation from the state space into the measurement space. Typically, the measurement space is smaller than the state space, i.e. $m < n$, which means that not all components of the object state are measurable. The residual of the actual measurement $\boldsymbol{z}_{k+1}$ and the predicted measurement $\hat{\boldsymbol{z}}_{k+1|k}$ is defined as 
\begin{equation}
\boldsymbol{\gamma}_{k+1} \coloneqq \boldsymbol{z}_{k+1} - \hat{\boldsymbol{z}}_{k+1|k}
\end{equation}
and is used in the innovation of the Kalman filter.
Then, the measurement $\boldsymbol{z}_{k+1}$ is taken into account during the update step yielding the posterior state estimation
\begin{align}
\hat{\boldsymbol{x}}_{k+1} &= \hat{\boldsymbol{x}}_{k+1|k} + \boldsymbol{K}_{k+1} \boldsymbol{\gamma}_{k+1}, \\
\boldsymbol{P}_{k+1} &= \boldsymbol{P}_{k+1|k} + 
\boldsymbol{K}_{k+1} \boldsymbol{S}_{k+1} \boldsymbol{K}_{k+1}^T ,
\end{align}
where the Kalman gain
\begin{equation}
\boldsymbol{K}_{k+1} = \boldsymbol{P}_{k+1|k} \boldsymbol{H}_{k+1}^T \boldsymbol{S}_{k+1}^{-1}
\end{equation}
models the impact of the process and measurement model uncertainties towards the posterior state estimation. For small values of $\boldsymbol{K}_{k+1}$, the posterior state estimation trusts more in the state prediction $\hat{\boldsymbol{x}}_{k+1|k}$, i.e., the process model, and, accordingly, for big values of $\boldsymbol{K}_{k+1}$, the posterior state estimation trusts more in the current measurement $\boldsymbol{z}_{k+1}$.

\subsection{Consistency of State Estimators}\label{section:ConsistencyStateEstimators}

For estimating static parameters, consistency is defined such that the estimated value must converge with increasing number of measurements to the true value.
For state estimation in dynamic systems, this consistency definition is not applicable due to the time-variant state. 
In~\cite{bar1988tracking}, practical consistency conditions of state estimators are defined as
\begin{align}
\mathbb{E} \left[ \boldsymbol{x}_{k} - \hat{\boldsymbol{x}}_{k}\right] &\coloneqq \mathbb{E} \left[ \tilde{\boldsymbol{x}}_{k}\right] \overset{!}{=} 0 , \label{eqn:consistency_cond_bias_free}\\
\mathbb{E} \left[ \tilde{\boldsymbol{x}}_{k} \tilde{\boldsymbol{x}}_{k}^T \right] &\overset{!}{=} \boldsymbol{P}_{k} ,\label{eqn:consistency_cond_cov_mse}
\end{align}
where~\eqref{eqn:consistency_cond_bias_free} depicts that the estimator should be unbiased and~\eqref{eqn:consistency_cond_cov_mse} describes that the mean square error should be equivalent to the estimated covariance matrix $\boldsymbol{P}_{k}$. For examining condition~\eqref{eqn:consistency_cond_cov_mse}, which implicitly include~\eqref{eqn:consistency_cond_bias_free}, the \textit{NEES}
\begin{equation}
\varepsilon_{\boldsymbol{x}_{k}} = \tilde{\boldsymbol{x}}_{k}^T \boldsymbol{P}_{k}^{-1} \tilde{\boldsymbol{x}}_{k}
\end{equation}
is used. The NEES follows a $\chi^2$ distribution with $n$ degrees of freedom (the dimension of the state space) if all assumptions of the Kalman filter are fulfilled. To check if the Kalman filter is consistent, the NEES must be in a certain confidence interval of the $\chi^2$ distribution. However, to perform the NEES, a ground truth is necessary, which is often not available.

For online applications the \textit{time-average NIS}~\cite{bar2001estimation}
\begin{equation} \label{eqn:time_avg_nis}
\bar{\varepsilon}_{\boldsymbol{\gamma}} = \frac{1}{K} \sum_{k=1}^{K}
\boldsymbol{\gamma}_{k}^T \boldsymbol{S}_{k}^{-1} \boldsymbol{\gamma}_{k}
\coloneqq \frac{1}{K} \sum_{k=1}^{K} \varepsilon_{\boldsymbol{\gamma}_{k}}
\end{equation}
is designed as a time-average value over a data window of size $K \in \mathbb{N}$ of the classical NIS $\varepsilon_{\boldsymbol{\gamma}_{k}}$, which is the Mahalanobis distance of the measurement residual $\boldsymbol{\gamma}_{k}$ with regard to the innovation covariance matrix $\boldsymbol{S}_{k}$. In fact, supposing ergodicity, if the  Kalman filter's assumptions are fulfilled, then $K \bar{\varepsilon}_{\boldsymbol{\gamma}}$ also follows a $\chi^2$ distribution with $K m$ degrees of freedom.

\section{Self-Assessment Method Using Subjective Logic} \label{section:self_assessing_method}

In this section, starting with our problem formulation, we present our proposed algorithm for self-assessing Kalman filtering using subjective logic and explain the respective steps in detail.

\subsection{Problem Formulation}

Given a Kalman filter, the goal of the proposed method is to realize an online self-assessment of the Kalman filter's performance. Therefore, the proposed method monitors the validity of the statistical assumptions of Kalman filtering in online applications. This objective is similar to the NIS or time-average NIS if multiple measurements are used. However, in contrast to the traditional NIS, we want to use a measure for consistency testing that is more significant in terms of statistical evidence.
Moreover, we want to generate a self-assessment online measure that also supplies an explicit certainty measure expressing the level of certainty about the statement. Hence, we can estimate the reliability of each sensor with respect to the filtering assumptions consisting of a self-assessment measure with an explicit certainty value of the measure.

Kalman filtering produces measurement prediction in terms of $\hat{\boldsymbol{z}}_{k+1|k} \in \mathbb{R}^m$ and $\boldsymbol{S}_{k+1} \in \mathbb{R}^{m \times m}$ for every time step $k \in \mathbb{N}$, see~\eqref{eqn:kf_meas_pred} and~\eqref{eqn:kf_meas_pred_cov}, respectively. Using subjective logic and the incoming measurements $\boldsymbol{z}_{k+1} \in \mathbb{R}^m$, the proposed method outputs a self-assessment online measure $\delta_k \in [0,1]$ and, additionally, a corresponding explicit uncertainty ${u_{\delta_k} \in [0,1]}$ in every time step based on the filtering measurement predictions and assumptions. As our method typically uses multiple measurements, we compare our measure to the time-average NIS for ensuring a fair comparison.

\subsection{Algorithm}

The key idea of the algorithm is to form a multinomial opinion of the correctness of the Kalman filter's assumptions with respect to the incoming measurements. Consequently, we compare the generated opinion with an 
ideal Gaussian opinion based on the filtering assumptions. This comparison leads to a DC which gives us a self-assessment measure and a corresponding explicit uncertainty of this measure. Algorithm~\ref{algo:SLConsistencyMeasure} portrays an overview of this procedure in order to determine a self-assessment online measure.
\begin{figure}[!t]
	\removelatexerror
	\begin{algorithm}[H]
		\caption{Self-assessing Kalman filter using subjective logic.}
		\label{algo:SLConsistencyMeasure}
		\small
		\begin{algorithmic}[1]
			\Require Random variable $X \in \mathbb{X} = \left\{x_1, \ldots, x_{n_X} \right\}$ with $n_X \in \mathbb{N}$ modeling the assumptions for a Gaussian distribution, initial opinion $\omega_X^0 = (\boldsymbol{b}_X^0,u_X^0,\boldsymbol{a}_X)$, reference opinion of the assumed Gaussian distribution  $\omega_X^G = (\boldsymbol{b}_X^G,u_X^G,\boldsymbol{a}_X)$, number of time steps $n \in \mathbb{N}$, window length $n_{st} \in \mathbb{N}$ for short-term opinion generation, step size $n_c \in \mathbb{N}$ with $n_c < n_{st}$ for long-term and short-term opinion comparison, threshold $\theta \in [0,1]$, trust discounting probability $p_d \in [0,1]$
			\Ensure Self-assessment online measure $\delta_k \in [0,1]$ of the correctness of the filtering assumptions with corresponding explicit uncertainty $u_{\delta_{k}} \in [0,1]$ for $k=0, \ldots, n$
			\vspace{5pt}
			
			\Procedure{SLConsistencyMeasure}{$X, \omega_X^0, \omega_X^G, n,$
				
			\qquad \qquad \qquad \qquad	\qquad \qquad \qquad \qquad 
			$n_{st}, n_c, \theta, p_d$}
			\State initialize $k \gets 0, i \gets 0, l \gets 0, \omega_X^{st_0} \gets \omega_X^0, \omega_X^{lt_0} \gets \omega_X^0,$ 
			
			$\delta_{0} \gets \text{DC} \left(\omega_X^{0}, \omega_X^{G}\right)$
			
			\While{$k < n-1$}
				
				\If{$k < n_{st}-1$}
				
					\State Obtain Kalman filter's measurement prediction 
				
					\qquad \quad $\hat{\boldsymbol{z}}_{k+1|k}, \boldsymbol{S}_{k+1}$ and incoming measurement $\boldsymbol{z}_{k+1}$

					\State $\left[\omega_X^{st_{k+1}}, \omega_X^{\boldsymbol{z}_{k+1}} \right]$ $\gets$ \textsc{UpdateOpinion}$\big(X, \omega_X^{st_{k}},$
				
					\qquad \qquad \qquad \qquad \qquad \qquad \qquad
					$\hat{\boldsymbol{z}}_{k+1|k}, \boldsymbol{S}_{k+1}, \boldsymbol{z}_{k+1}\big)$
					\State $\omega_X^{k+1} \gets \omega_X^{st_{k+1}}$
					
					\State $\delta_{k+1} \gets \text{DC} \left(\omega_X^{k+1}, \omega_X^{G}\right)$
					
					\State $u_{\delta_{k+1}} \gets u_X^{k+1}$
					
					\State $k \gets k+1$

				\Else
			
					\For{$j = 0, \ldots, n_c-1$}
						
						\State Obtain Kalman filter's measurement prediction 
						
						\qquad \qquad \enskip $\hat{\boldsymbol{z}}_{k+1|k}, \boldsymbol{S}_{k+1}$ and incoming measurement $\boldsymbol{z}_{k+1}$

						\State $\left[\omega_X^{st_{k+1}}, \omega_X^{\boldsymbol{z}_{k+1}} \right]$ $\gets$ \textsc{UpdateOpinion}$\big(X, \omega_X^{st_{k}},$
						
						\qquad \qquad \qquad \qquad \qquad \qquad \qquad \quad \;
						$\hat{\boldsymbol{z}}_{k+1|k}, \boldsymbol{S}_{k+1}, \boldsymbol{z}_{k+1}\big)$
						
						\State $\omega_X^{st_{k+1}} \gets \omega_X^{st_{k+1}} \ominus \omega_X^{\boldsymbol{z}_{k-n_{st}+1}}$
						
						\State $\omega_X^{lt_i} \gets \omega_X^{lt_i} \oplus \omega_X^{\boldsymbol{z}_{k-n_{st}+1}}$
						
						\State $\omega_X^{k+1} \gets \omega_X^{st_{k+1}} \oplus \omega_X^{lt_i}$
						
						\State $\delta_{k+1} \gets \text{DC} \left(\omega_X^{k+1}, \omega_X^{G}\right)$
						
						\State $u_{\delta_{k+1}} \gets u_X^{k+1}$
						
						\State $k \gets k+1$, $l \gets l+1$
						
					\EndFor
					
					\If{$\text{DC} \left(\omega_X^{lt_i}, \omega_X^{st_{k+1}}\right) > \theta$ and $l \geq n_{st}$}
						\State $\omega_X^{lt_{i+1}} \gets  \omega_X^{0}$, $l \gets  0$
					\ElsIf{$l \geq n_{st}$}
						\State $\omega_X^{lt_{i+1}} \gets \omega_X^{lt_{i}}$
						\State $\omega_X^{lt_{i+1}} \gets \text{TD}\left(\omega_X^{lt_{i+1}},  p_d \right)$
					\Else
						\State $\omega_X^{lt_{i+1}} \gets \omega_X^{lt_{i}}$
					\EndIf
					\State $i \gets i+1$
				
				\EndIf
				
			\EndWhile
			
			\State \textbf{return} $\boldsymbol{\delta} = \left[ \delta_0, \ldots, \delta_{n} \right], \boldsymbol{u}_{\delta} = \left[ u_{\delta_0}, \ldots, u_{\delta_{n}} \right]$
			
			\EndProcedure
		\end{algorithmic} 
	\end{algorithm}
\end{figure}
As input, our proposed method needs a random variable $X \in \mathbb{X} = \left\{x_1, \ldots, x_{n_X} \right\}$, $n_X \in \mathbb{N}$, which models the Gaussian distribution assumptions of Kalman filtering. This is implemented by discretizing the assumed Gaussian distribution in $n_X$ bins in order to use the evidence of our samples, i.e., the incoming measurements, in a supported subjective logic manner.
Moreover, the initial opinion $\omega_X^0 = (\boldsymbol{b}_X^0,u_X^0,\boldsymbol{a}_X)$ of the correctness of the Kalman filter's assumptions is constituted as vacuous opinions, i.e., $u_X^0=1$ and $\boldsymbol{b}_X^0(x) = 0 \; \forall x \in \mathbb{X}$. In addition, a reference opinion $\omega_X^G = (\boldsymbol{b}_X^G,u_X^G,\boldsymbol{a}_X)$ of the assumed Gaussian distribution is featured as a dogmatic opinion, i.e., $u_X^0=0$. Further, the number of time steps $n \in \mathbb{N}$ is specified. To be able to correctly monitor drifts and jumps in the ground truth noise parameters, we define the window length $n_{st} \in \mathbb{N}$ for the short-term opinion generation, the number of time steps $n_c \in \mathbb{N}$ with $n_c < n_{st}$ for short-term and long-term opinion comparison, and a threshold $\theta \in [0,1]$ for the corresponding comparison using the DC. For modeling the degradation of information for time-varying parameters over time, trust discounting is applied with respect to the probability $p_d \in [0,1]$. To neglect this aspect, the discount probability can be chosen to $p_d = 1$.

After the initialization step, the first $n_{st}$ time steps are used to generate a short-term opinion about the correctness of the filter assumptions with respect to the incoming measurements. One important component, in doing so, is the procedure of updating the previous opinion with the incoming measurement. This procedure is displayed in Algorithm~\ref{algo:UpdateOpinion}.
\begin{figure}[!t]
	\removelatexerror
	\begin{algorithm}[H]
		\caption{Update opinion with incoming measurement.}
		\label{algo:UpdateOpinion}
		\small
		\begin{algorithmic}[1]
			\Require Random variable $X \in \mathbb{X}$ with cardinality $n_X = |\mathbb{X}| \geq 2$,
			opinion $\omega_X$ over $X$, Kalman filter's measurement prediction $\hat{\boldsymbol{z}} \in \mathbb{R}^{m}$ and covariance matrix $\boldsymbol{S} \in \mathbb{R}^{m \times m}$ with $m \in \mathbb{N}$, measurement  $\boldsymbol{z} \in \mathbb{R}^{m}$
			\Ensure Updated opinion $\bar{\omega}_X$ over $X$, generated opinion $\omega_X^{\tilde{\boldsymbol{z}}}$ over $X$ with respect to the transformed measurement $\tilde{\boldsymbol{z}}$
			\vspace{5pt}
			
			\Procedure{UpdateOpinion}{$X, \omega_X, \hat{\boldsymbol{z}}, \boldsymbol{S}, \boldsymbol{z}$}
			
			\State $\tilde{\boldsymbol{z}} \gets \boldsymbol{S}^{-1/2} \left( \boldsymbol{z} - \hat{\boldsymbol{z}} \right)$
			\State Generate opinion $\omega_X^{\tilde{\boldsymbol{z}}}$ over $X$ with respect to 
			
			the transformed measurement $\tilde{\boldsymbol{z}}$
			\State $\bar{\omega}_X \gets \omega_X \oplus \omega_X^{\tilde{\boldsymbol{z}}}$
			\State \textbf{return} $\bar{\omega}_X$, $\omega_X^{\tilde{\boldsymbol{z}}}$
			
			\EndProcedure
		\end{algorithmic} 
	\end{algorithm}
\end{figure}
In addition to the previous short-term opinion $\omega_X$, this procedure needs the Kalman filter's measurement prediction $\hat{\boldsymbol{z}} \in \mathbb{R}^{m}$, the covariance matrix $\boldsymbol{S} \in \mathbb{R}^{m \times m}$, and the incoming measurement $\boldsymbol{z} \in \mathbb{R}^{m}$ as input parameters. Then, the incoming measurement is mapped to a standard normal distribution based on the measurement prediction of Kalman filtering. The notation $\boldsymbol{S}^{-1/2}$ denotes the square root of the inverse of covariance matrix $\boldsymbol{S}$, which can be obtained using, e.g., the Cholesky factorization. For further details, please refer to~\cite{bar2001estimation}. Consequently, the transformed measurement $\tilde{\boldsymbol{z}}$ is assigned to a certain event $x_i$ of $X$, $i \in \{1, \ldots, n_Z\}$ such that a resulting opinion $\omega_X^{\tilde{\boldsymbol{z}}}$ with respect to the measurement is generated. This opinion is fused with the previous opinion to generate the updated opinion $\bar{\omega}_X$. To conclude, the procedure returns the updated opinion and the generated opinion with respect to the measurement.

Continuing the procedure of Algorithm~\ref{algo:SLConsistencyMeasure} and after processing the first $n_{st}$ time steps, we update the short-term and long-term opinion $n_c$ times while calculating the self-assessment measure in each time step. After $n_c$ time steps, we compare the short-term and long-term performance represented by opinions using the DC such that we are able to react quickly to sudden noise parameter changes, which are noticeable in the short-term opinion. This is implemented such that, if the opinions match, i.e., the DC is smaller than a threshold, new opinions will be continuously merged with previous opinions, which is based on more statistical data and, hence, show less statistical uncertainty. On the downside, if the opinions do not match, i.e., the DC exceeds a certain threshold, the previous long-term opinion will be discarded. This procedure continues until time step $n$ and is able to output the self-assessment online measure $\delta_k$ and the corresponding uncertainty $u_{\delta_k}$ in every time step.

\section{Simulation Results}\label{section:experiments}

This section evaluates our proposed self-assessment method through simulated data. On the one hand, jumps and drifts in the ground truth measurement noise parameters are examined and, on the other hand, changes in the process model for the generation of ground truth data are evaluated. 

For the following simulations, we consider a single-target multi-sensor simulation setup with two sensors measuring the position in one dimension of a single object in each time step. The two sensors are modeled to be equal in terms of Kalman filter's assumptions, i.e., the measurement noise is assumed to be $w_{k} \sim \mathcal{N}\left( 0, \sigma_w\right)$ with constant variance $\sigma_w$ for both sensors. Moreover, we assume a constant velocity model with process noise $v_{k} \sim \mathcal{N}\left( 0, \sigma_v\right)$. Further, it is assumed that $\sigma_w = \sigma_v$, which should describe the fact that we do not have prior knowledge about the noise parameters. The parameters of our proposed method are chosen in the following way. We choose $n_{st} = 35$ to incorporate enough evidence to form a reliable short-term opinion, $n_c = 1$ to be able to react quickly on parameter jumps, $\theta = 0.25$ to define a threshold for the comparison of subjective opinions, and $p_d = 0.99$ to apply trust discounting.

\subsection{Jumps in Measurement Noise}

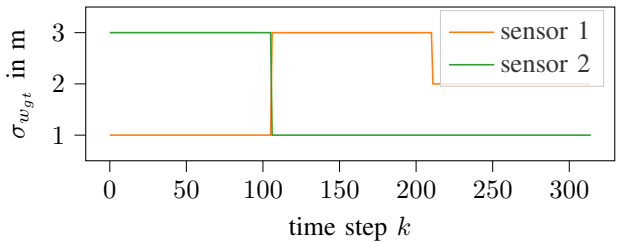
\begin{figure}[b]
	\centering
\begin{tikzpicture}

\definecolor{color0}{rgb}{1,0.498039215686275,0.0549019607843137}
\definecolor{color1}{rgb}{0.172549019607843,0.627450980392157,0.172549019607843}

\begin{axis}[
height=0.2\textwidth,
width=0.475\textwidth,
legend cell align={left},
legend style={fill opacity=0.8, draw opacity=1, text opacity=1, draw=white!80!black, at={(0.98,0.98)},anchor=north east},
tick align=outside,
tick pos=left,
x grid style={white!69.0196078431373!black},
xlabel={time step $k$},
xmin=-15.7, xmax=329.7,
xtick style={color=black},
y grid style={white!69.0196078431373!black},
ylabel={$\sigma_{w_{gt}}$ in m},
ymin=0.5, ymax=3.5,
ytick style={color=black}
]
\addplot [semithick, color0]
table {%
0 1
1 1
2 1
3 1
4 1
5 1
6 1
7 1
8 1
9 1
10 1
11 1
12 1
13 1
14 1
15 1
16 1
17 1
18 1
19 1
20 1
21 1
22 1
23 1
24 1
25 1
26 1
27 1
28 1
29 1
30 1
31 1
32 1
33 1
34 1
35 1
36 1
37 1
38 1
39 1
40 1
41 1
42 1
43 1
44 1
45 1
46 1
47 1
48 1
49 1
50 1
51 1
52 1
53 1
54 1
55 1
56 1
57 1
58 1
59 1
60 1
61 1
62 1
63 1
64 1
65 1
66 1
67 1
68 1
69 1
70 1
71 1
72 1
73 1
74 1
75 1
76 1
77 1
78 1
79 1
80 1
81 1
82 1
83 1
84 1
85 1
86 1
87 1
88 1
89 1
90 1
91 1
92 1
93 1
94 1
95 1
96 1
97 1
98 1
99 1
100 1
101 1
102 1
103 1
104 1
105 1
106 3
107 3
108 3
109 3
110 3
111 3
112 3
113 3
114 3
115 3
116 3
117 3
118 3
119 3
120 3
121 3
122 3
123 3
124 3
125 3
126 3
127 3
128 3
129 3
130 3
131 3
132 3
133 3
134 3
135 3
136 3
137 3
138 3
139 3
140 3
141 3
142 3
143 3
144 3
145 3
146 3
147 3
148 3
149 3
150 3
151 3
152 3
153 3
154 3
155 3
156 3
157 3
158 3
159 3
160 3
161 3
162 3
163 3
164 3
165 3
166 3
167 3
168 3
169 3
170 3
171 3
172 3
173 3
174 3
175 3
176 3
177 3
178 3
179 3
180 3
181 3
182 3
183 3
184 3
185 3
186 3
187 3
188 3
189 3
190 3
191 3
192 3
193 3
194 3
195 3
196 3
197 3
198 3
199 3
200 3
201 3
202 3
203 3
204 3
205 3
206 3
207 3
208 3
209 3
210 3
211 2
212 2
213 2
214 2
215 2
216 2
217 2
218 2
219 2
220 2
221 2
222 2
223 2
224 2
225 2
226 2
227 2
228 2
229 2
230 2
231 2
232 2
233 2
234 2
235 2
236 2
237 2
238 2
239 2
240 2
241 2
242 2
243 2
244 2
245 2
246 2
247 2
248 2
249 2
250 2
251 2
252 2
253 2
254 2
255 2
256 2
257 2
258 2
259 2
260 2
261 2
262 2
263 2
264 2
265 2
266 2
267 2
268 2
269 2
270 2
271 2
272 2
273 2
274 2
275 2
276 2
277 2
278 2
279 2
280 2
281 2
282 2
283 2
284 2
285 2
286 2
287 2
288 2
289 2
290 2
291 2
292 2
293 2
294 2
295 2
296 2
297 2
298 2
299 2
300 2
301 2
302 2
303 2
304 2
305 2
306 2
307 2
308 2
309 2
310 2
311 2
312 2
313 2
314 2
};
\addlegendentry{sensor 1}
\addplot [semithick, color1]
table {%
0 3
1 3
2 3
3 3
4 3
5 3
6 3
7 3
8 3
9 3
10 3
11 3
12 3
13 3
14 3
15 3
16 3
17 3
18 3
19 3
20 3
21 3
22 3
23 3
24 3
25 3
26 3
27 3
28 3
29 3
30 3
31 3
32 3
33 3
34 3
35 3
36 3
37 3
38 3
39 3
40 3
41 3
42 3
43 3
44 3
45 3
46 3
47 3
48 3
49 3
50 3
51 3
52 3
53 3
54 3
55 3
56 3
57 3
58 3
59 3
60 3
61 3
62 3
63 3
64 3
65 3
66 3
67 3
68 3
69 3
70 3
71 3
72 3
73 3
74 3
75 3
76 3
77 3
78 3
79 3
80 3
81 3
82 3
83 3
84 3
85 3
86 3
87 3
88 3
89 3
90 3
91 3
92 3
93 3
94 3
95 3
96 3
97 3
98 3
99 3
100 3
101 3
102 3
103 3
104 3
105 3
106 1
107 1
108 1
109 1
110 1
111 1
112 1
113 1
114 1
115 1
116 1
117 1
118 1
119 1
120 1
121 1
122 1
123 1
124 1
125 1
126 1
127 1
128 1
129 1
130 1
131 1
132 1
133 1
134 1
135 1
136 1
137 1
138 1
139 1
140 1
141 1
142 1
143 1
144 1
145 1
146 1
147 1
148 1
149 1
150 1
151 1
152 1
153 1
154 1
155 1
156 1
157 1
158 1
159 1
160 1
161 1
162 1
163 1
164 1
165 1
166 1
167 1
168 1
169 1
170 1
171 1
172 1
173 1
174 1
175 1
176 1
177 1
178 1
179 1
180 1
181 1
182 1
183 1
184 1
185 1
186 1
187 1
188 1
189 1
190 1
191 1
192 1
193 1
194 1
195 1
196 1
197 1
198 1
199 1
200 1
201 1
202 1
203 1
204 1
205 1
206 1
207 1
208 1
209 1
210 1
211 1
212 1
213 1
214 1
215 1
216 1
217 1
218 1
219 1
220 1
221 1
222 1
223 1
224 1
225 1
226 1
227 1
228 1
229 1
230 1
231 1
232 1
233 1
234 1
235 1
236 1
237 1
238 1
239 1
240 1
241 1
242 1
243 1
244 1
245 1
246 1
247 1
248 1
249 1
250 1
251 1
252 1
253 1
254 1
255 1
256 1
257 1
258 1
259 1
260 1
261 1
262 1
263 1
264 1
265 1
266 1
267 1
268 1
269 1
270 1
271 1
272 1
273 1
274 1
275 1
276 1
277 1
278 1
279 1
280 1
281 1
282 1
283 1
284 1
285 1
286 1
287 1
288 1
289 1
290 1
291 1
292 1
293 1
294 1
295 1
296 1
297 1
298 1
299 1
300 1
301 1
302 1
303 1
304 1
305 1
306 1
307 1
308 1
309 1
310 1
311 1
312 1
313 1
314 1
};
\addlegendentry{sensor 2}
\end{axis}

\end{tikzpicture}
	\caption{Jumps in the ground truth measurement noise of the sensor data.\label{figure:exp01_meas_noise}}
\end{figure}

\begin{figure}
	\centering
	\input{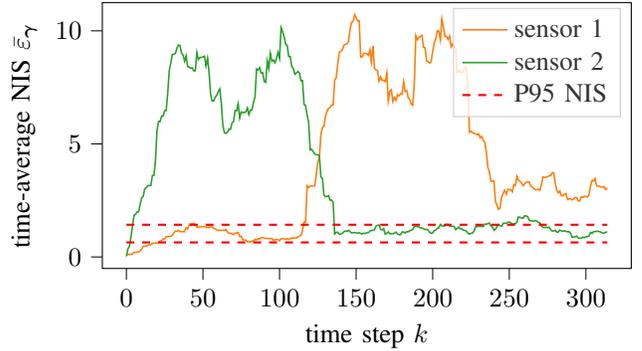}
	\caption{Time-average NIS of the experiment with jumps in the ground truth measurement noise.\label{figure:exp01_time_avg_nis_gauss}}
\end{figure}

\begin{figure}
	\centering
	\subfloat[Self-assessment measure.\label{figure:exp01_dc_gauss}]{%
		\centering
		\input{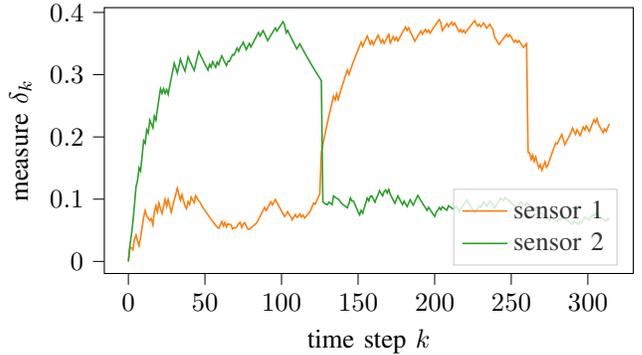}}
	\\
	\subfloat[Uncertainty of the self-assessment measure.\label{figure:exp01_uncertainty_gauss}]{%
		\centering
		\input{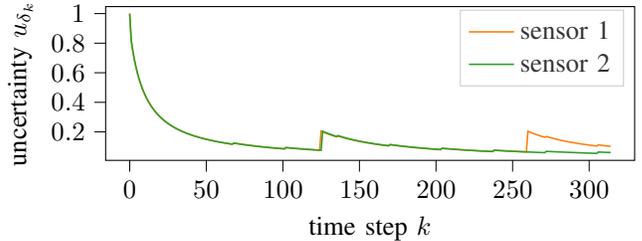}}
	\caption{Results of the experiment with jumps in the ground truth measurement noise for our self-assessment method based on subjective logic.\label{figure:exp01_our_result}}
\end{figure}

We first consider jumps in our simulated ground truth measurement noise $\sigma_{w_{gt}}$. The progress of the ground truth measurement noise of our simulated sensors is illustrated in Fig.~\ref{figure:exp01_meas_noise},
where two jumps are located at time step $105$ and $210$. The underlying process model for the ground truth data generation and for the Kalman filter is a constant velocity model. 
The results of the first simulation scenario in terms of the time-average NIS are shown in Fig.~\ref{figure:exp01_time_avg_nis_gauss}.
The $95 \%$ confidence interval of the time-average NIS is displayed as reference. It can be seen that the Kalman filter's assumptions are violated by these jumps in the simulated measurement noise during the corresponding time sections.
With our proposed method, we obtain a self-assessment measure in Fig.~\subref{figure:exp01_dc_gauss} and the corresponding uncertainty in Fig.~\subref{figure:exp01_uncertainty_gauss}.
We obtain similar results with the subjective logic-based measure as the time-average NIS in Fig.~\ref{figure:exp01_time_avg_nis_gauss}. However, the scales of the two measures are different. The time-average NIS is given as the support of a $\chi^2$ distribution, i.e., the interval $[0,\infty)$. In contrast, the subjective logic self-assessment measure is given as the DC between two opinions, i.e., as a normalized value in $\left[0,1\right]$. Compared to the time-average NIS, our proposed self-assessment measure shows sharp edges when recognizing jumps and keeps the level of the measure more constant during the jumps. Particularly, the time-average NIS shows small collapses during the jumps.
In addition, the peaks in the uncertainty in Fig.~\subref{figure:exp01_uncertainty_gauss} support the conclusion that our proposed method has recognized the jumps and has consequently discarded the long-term history. Compared to the ground truth, these jumps are detected with small delays as well as for the time-average NIS. However this is plausible because in order to recognize jumps, and to be certain about it, a certain amount of statistical data has to be collected.

\subsection{Drift in Measurement Noise}

\begin{figure}
	\centering
\begin{tikzpicture}

\definecolor{color0}{rgb}{1,0.498039215686275,0.0549019607843137}
\definecolor{color1}{rgb}{0.172549019607843,0.627450980392157,0.172549019607843}

\begin{axis}[
height=0.2\textwidth,
width=0.475\textwidth,
legend pos=north west,
legend cell align={left},
legend style={fill opacity=0.8, draw opacity=1, text opacity=1, draw=white!80!black},
tick align=outside,
tick pos=left,
x grid style={white!69.0196078431373!black},
xlabel={time step $k$},
xmin=-6.7, xmax=140.7,
xtick style={color=black},
y grid style={white!69.0196078431373!black},
ylabel={$\sigma_{w_{gt}}$ in m},
ymin=0.5, ymax=3.5,
ytick style={color=black}
]
\addplot [semithick, color0]
table {%
0 1.01481481481481
1 1.02962962962963
2 1.04444444444444
3 1.05925925925926
4 1.07407407407407
5 1.08888888888889
6 1.1037037037037
7 1.11851851851852
8 1.13333333333333
9 1.14814814814815
10 1.16296296296296
11 1.17777777777778
12 1.19259259259259
13 1.20740740740741
14 1.22222222222222
15 1.23703703703704
16 1.25185185185185
17 1.26666666666667
18 1.28148148148148
19 1.2962962962963
20 1.31111111111111
21 1.32592592592593
22 1.34074074074074
23 1.35555555555556
24 1.37037037037037
25 1.38518518518519
26 1.4
27 1.41481481481482
28 1.42962962962963
29 1.44444444444445
30 1.45925925925926
31 1.47407407407407
32 1.48888888888889
33 1.5037037037037
34 1.51851851851852
35 1.53333333333333
36 1.54814814814815
37 1.56296296296296
38 1.57777777777778
39 1.59259259259259
40 1.60740740740741
41 1.62222222222222
42 1.63703703703704
43 1.65185185185185
44 1.66666666666667
45 1.68148148148148
46 1.6962962962963
47 1.71111111111111
48 1.72592592592593
49 1.74074074074074
50 1.75555555555556
51 1.77037037037037
52 1.78518518518519
53 1.8
54 1.81481481481482
55 1.82962962962963
56 1.84444444444445
57 1.85925925925926
58 1.87407407407408
59 1.88888888888889
60 1.90370370370371
61 1.91851851851852
62 1.93333333333333
63 1.94814814814815
64 1.96296296296296
65 1.97777777777778
66 1.99259259259259
67 2.00740740740741
68 2.02222222222222
69 2.03703703703704
70 2.05185185185185
71 2.06666666666667
72 2.08148148148148
73 2.0962962962963
74 2.11111111111111
75 2.12592592592593
76 2.14074074074074
77 2.15555555555556
78 2.17037037037037
79 2.18518518518518
80 2.2
81 2.21481481481481
82 2.22962962962963
83 2.24444444444444
84 2.25925925925926
85 2.27407407407407
86 2.28888888888889
87 2.3037037037037
88 2.31851851851852
89 2.33333333333333
90 2.34814814814814
91 2.36296296296296
92 2.37777777777777
93 2.39259259259259
94 2.4074074074074
95 2.42222222222222
96 2.43703703703703
97 2.45185185185185
98 2.46666666666666
99 2.48148148148148
100 2.49629629629629
101 2.51111111111111
102 2.52592592592592
103 2.54074074074073
104 2.55555555555555
105 2.57037037037036
106 2.58518518518518
107 2.59999999999999
108 2.61481481481481
109 2.62962962962962
110 2.64444444444444
111 2.65925925925925
112 2.67407407407407
113 2.68888888888888
114 2.7037037037037
115 2.71851851851851
116 2.73333333333332
117 2.74814814814814
118 2.76296296296295
119 2.77777777777777
120 2.79259259259258
121 2.8074074074074
122 2.82222222222221
123 2.83703703703703
124 2.85185185185184
125 2.86666666666666
126 2.88148148148147
127 2.89629629629629
128 2.9111111111111
129 2.92592592592591
130 2.94074074074073
131 2.95555555555554
132 2.97037037037036
133 2.98518518518517
134 2.99999999999999
};
\addlegendentry{sensor 1}
\addplot [semithick, color1]
table {%
0 1
1 1
2 1
3 1
4 1
5 1
6 1
7 1
8 1
9 1
10 1
11 1
12 1
13 1
14 1
15 1
16 1
17 1
18 1
19 1
20 1
21 1
22 1
23 1
24 1
25 1
26 1
27 1
28 1
29 1
30 1
31 1
32 1
33 1
34 1
35 1
36 1
37 1
38 1
39 1
40 1
41 1
42 1
43 1
44 1
45 1
46 1
47 1
48 1
49 1
50 1
51 1
52 1
53 1
54 1
55 1
56 1
57 1
58 1
59 1
60 1
61 1
62 1
63 1
64 1
65 1
66 1
67 1
68 1
69 1
70 1
71 1
72 1
73 1
74 1
75 1
76 1
77 1
78 1
79 1
80 1
81 1
82 1
83 1
84 1
85 1
86 1
87 1
88 1
89 1
90 1
91 1
92 1
93 1
94 1
95 1
96 1
97 1
98 1
99 1
100 1
101 1
102 1
103 1
104 1
105 1
106 1
107 1
108 1
109 1
110 1
111 1
112 1
113 1
114 1
115 1
116 1
117 1
118 1
119 1
120 1
121 1
122 1
123 1
124 1
125 1
126 1
127 1
128 1
129 1
130 1
131 1
132 1
133 1
134 1
};
\addlegendentry{sensor 2}
\end{axis}

\end{tikzpicture}
	\caption{Drift in the ground truth measurement noise of the sensor data.\label{figure:exp02_meas_noise}}
\end{figure}
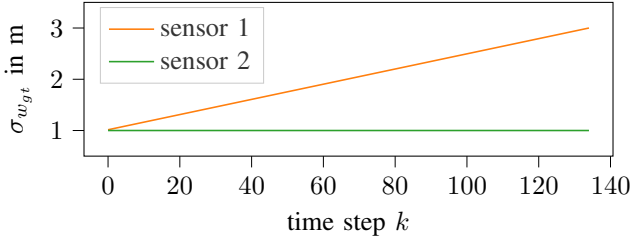

\begin{figure}
	\centering
	\input{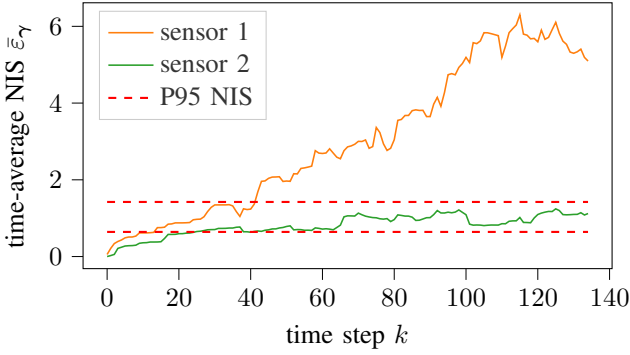}
	\caption{Time-average NIS of the experiment with a drift in the ground truth measurement noise.\label{figure:exp02_time_avg_nis_gauss}}
\end{figure}

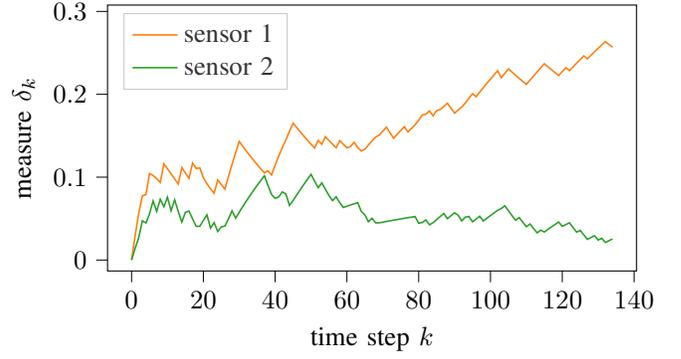
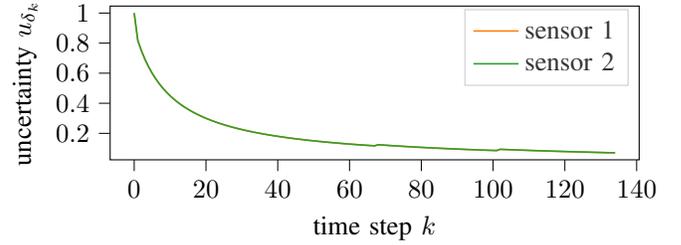
\begin{figure}
	\centering
	\subfloat[Self-assessment measure.\label{figure:exp02_dc_gauss}]{%
\begin{tikzpicture}

\definecolor{color0}{rgb}{1,0.498039215686275,0.0549019607843137}
\definecolor{color1}{rgb}{0.172549019607843,0.627450980392157,0.172549019607843}

\begin{axis}[
height=0.28225\textwidth,
width=0.475\textwidth,
legend cell align={left},
legend style={fill opacity=0.8, draw opacity=1, text opacity=1, at={(0.03,0.97)}, anchor=north west, draw=white!80!black},
tick align=outside,
tick pos=left,
x grid style={white!69.0196078431373!black},
xlabel={time step $k$},
xmin=-6.7, xmax=140.7,
xtick style={color=black},
y grid style={white!69.0196078431373!black},
ylabel={measure $\delta_{k}$},
ymin=-0.0131757797808791, ymax=0.3076691375398461,
ytick style={color=black},
]
\addplot [semithick, color0]
table {%
0 0
1 0.0298207768987867
2 0.0563799063242687
3 0.0775260960898269
4 0.0789374221873465
5 0.104352435725141
6 0.102049219169887
7 0.0982963354903909
8 0.093606236376088
9 0.116030301924874
10 0.110207990459511
11 0.104120617126138
12 0.0979110059945177
13 0.0916793958978527
14 0.111537522115415
15 0.105008902123027
16 0.0985994877147961
17 0.11702837084144
18 0.110474298177953
19 0.111307118142451
20 0.0995047091216397
21 0.0919175699638059
22 0.08611514383097
23 0.0805039197525314
24 0.0967070663650499
25 0.0910658837594991
26 0.085614031846198
27 0.100798832519983
28 0.11542185384137
29 0.129508514762601
30 0.143083427337996
31 0.137134000777851
32 0.131376851277913
33 0.125804767323315
34 0.120410641883923
35 0.115187520678341
36 0.110128636502789
37 0.105227432813971
38 0.107692187766843
39 0.102728531362998
40 0.114315168398025
41 0.125541311877093
42 0.135499384590919
43 0.145112549456256
44 0.155332936483028
45 0.165250726796371
46 0.159891252788955
47 0.154680641226818
48 0.149613184410917
49 0.144683428837502
50 0.13988616489472
51 0.135216416520093
52 0.144457152064039
53 0.13984611076082
54 0.148781261986779
55 0.144229538224527
56 0.139792117381092
57 0.135464923599481
58 0.144003575946525
59 0.139728196330186
60 0.135555564222843
61 0.136888079350083
62 0.142144904906664
63 0.135629333281935
64 0.131694064403552
65 0.133716916714737
66 0.138766780438167
67 0.143702043036907
68 0.148526483330919
69 0.15063544642455
70 0.155581729608344
71 0.160413667461644
72 0.153593988358041
73 0.146926259898273
74 0.151690711493157
75 0.156349776091026
76 0.160906847756341
77 0.154443844832724
78 0.158922767012755
79 0.163306661525264
80 0.168514509033603
81 0.175166587147493
82 0.175917029629839
83 0.17994909751952
84 0.174007606040441
85 0.180338805670999
86 0.181610082625898
87 0.18542747614133
88 0.189170688218092
89 0.183171122621682
90 0.177284720216534
91 0.180997568401665
92 0.184640937916485
93 0.189516864208505
94 0.195182511173375
95 0.200745171068064
96 0.197144008537701
97 0.202589491766467
98 0.2079385541723
99 0.213193708018549
100 0.218357380128069
101 0.223431915428644
102 0.228419580326826
103 0.219947094054863
104 0.225354680475749
105 0.230663535591144
106 0.226780125617142
107 0.222965610191134
108 0.219218195913642
109 0.215536150082586
110 0.21191779820087
111 0.217065506930938
112 0.222124808741919
113 0.227097940350997
114 0.231987064505639
115 0.236794272973818
116 0.233177698898041
117 0.2296202039853
118 0.226120367142193
119 0.222676811922005
120 0.227364598110852
121 0.231977562604123
122 0.228568547716203
123 0.233099355638445
124 0.237559510750841
125 0.241950640833659
126 0.24627432447406
127 0.242885048192964
128 0.247136052176273
129 0.251323211362221
130 0.255447943832822
131 0.259511626305693
132 0.263515595617582
133 0.260147612186425
134 0.256828066534237
};
\addlegendentry{sensor 1}
\addplot [semithick, color1]
table {%
0 0
1 0.0143269080244944
2 0.0270868104838097
3 0.0473790710886162
4 0.0446991630578454
5 0.0558285756244077
6 0.0713444060610997
7 0.0588169443475077
8 0.0733432605242481
9 0.0643728955899479
10 0.0757968113066128
11 0.05955466223011
12 0.0725942141941059
13 0.0587082642846239
14 0.0457670857217957
15 0.0575878202651503
16 0.0591118317160306
17 0.0496341207015333
18 0.0406516200196122
19 0.0406504013932246
20 0.0477123931086682
21 0.0544914203556288
22 0.0385373936167732
23 0.0452059339791099
24 0.0345103839286678
25 0.0400207087298613
26 0.0409046034393859
27 0.0498368789088854
28 0.0592905936928389
29 0.0504989626245113
30 0.05771762096386
31 0.0646786544276072
32 0.0713938751722091
33 0.0778745952812783
34 0.0841316171005826
35 0.090175233530227
36 0.0960152354683502
37 0.101660924293203
38 0.0901940444603647
39 0.0790847681257859
40 0.0744075650713655
41 0.076231002070709
42 0.082224720312954
43 0.0797065383544329
44 0.065931609184427
45 0.0717353981155233
46 0.0784279141840385
47 0.0849281283445584
48 0.0912438256927568
49 0.0973824075834635
50 0.103350911803966
51 0.095181293931512
52 0.0872286934112053
53 0.0930903982753412
54 0.085370037735748
55 0.0778477512201317
56 0.0716861775633363
57 0.0762892557740783
58 0.0691581131261215
59 0.0633878307103696
60 0.0649046818677387
61 0.0663844636191229
62 0.0678284880147197
63 0.0692380077772272
64 0.0587442413867074
65 0.0544544640845032
66 0.0462149469119162
67 0.0504947970946924
68 0.044888833354853
69 0.0448582382045256
70 0.0457006330885472
71 0.0465234243874642
72 0.04732727577634
73 0.0481128220377165
74 0.0488806705642782
75 0.0496314027733015
76 0.0503655754383961
77 0.0510837219437096
78 0.0517863534654557
79 0.0524739600853194
80 0.0443348423340674
81 0.0453952197752344
82 0.0484078445334922
83 0.0425241496937586
84 0.0450460748427139
85 0.048808681389279
86 0.0524971523827974
87 0.0561136257747287
88 0.0498961088777338
89 0.0534680247758991
90 0.0569721486948501
91 0.0539951660753521
92 0.0472457546218486
93 0.0517475934636506
94 0.0525749149942827
95 0.0461025936274617
96 0.0494835426741319
97 0.0528042441844736
98 0.0471625587204799
99 0.0504453083399063
100 0.0536710510484036
101 0.0568412437742239
102 0.0599572948378432
103 0.0621346710174757
104 0.0654432707850569
105 0.0595127635612098
106 0.0536889482615215
107 0.0479690059982172
108 0.0512854006641483
109 0.0456867271283482
110 0.0401853846839566
111 0.043482870006076
112 0.0380943482287721
113 0.032797187334478
114 0.0360734361088786
115 0.0338215620739434
116 0.0369122527704191
117 0.0399521480373411
118 0.0429424786901574
119 0.0458844365250866
120 0.0407027825342882
121 0.0427405952828397
122 0.0449529346351616
123 0.0392439083907958
124 0.0336235732646519
125 0.0358549184907184
126 0.0303464864490254
127 0.0249217199747663
128 0.0271681211460907
129 0.0293808715328982
130 0.0240841155158639
131 0.0262869287392345
132 0.0210903587482702
133 0.0233208161193074
134 0.0254437157588434
};
\addlegendentry{sensor 2}
\end{axis}

\end{tikzpicture}}
	\\
	\subfloat[Uncertainty of the self-assessment measure.\label{figure:exp02_uncertainty_gauss}]{%
		\centering
\begin{tikzpicture}

\definecolor{color0}{rgb}{1,0.498039215686275,0.0549019607843137}
\definecolor{color1}{rgb}{0.172549019607843,0.627450980392157,0.172549019607843}

\begin{axis}[
height=0.2\textwidth,
width=0.475\textwidth,
legend cell align={left},
legend style={fill opacity=0.8, draw opacity=1, text opacity=1, draw=white!80!black},
tick align=outside,
tick pos=left,
x grid style={white!69.0196078431373!black},
xlabel={time step $k$},
xmin=-6.7, xmax=140.7,
xtick style={color=black},
y grid style={white!69.0196078431373!black},
ylabel={uncertainty $u_{\delta_{k}}$},
ymin=0.0238012806217184, ymax=1.04648565330373,
ytick style={color=black},
]
\addplot [semithick, color0]
table {%
0 1
1 0.818181818181818
2 0.75
3 0.692307692307692
4 0.642857142857143
5 0.6
6 0.5625
7 0.529411764705882
8 0.5
9 0.473684210526316
10 0.45
11 0.428571428571429
12 0.409090909090909
13 0.391304347826087
14 0.375
15 0.36
16 0.346153846153846
17 0.333333333333333
18 0.321428571428571
19 0.310344827586207
20 0.3
21 0.290322580645161
22 0.28125
23 0.272727272727273
24 0.264705882352941
25 0.257142857142857
26 0.25
27 0.243243243243243
28 0.236842105263158
29 0.230769230769231
30 0.225
31 0.219512195121951
32 0.214285714285714
33 0.209302325581395
34 0.204545454545455
35 0.2
36 0.195652173913043
37 0.191489361702128
38 0.1875
39 0.183673469387755
40 0.18
41 0.176470588235294
42 0.173076923076923
43 0.169811320754717
44 0.166666666666667
45 0.163636363636364
46 0.160714285714286
47 0.157894736842105
48 0.155172413793103
49 0.152542372881356
50 0.15
51 0.147540983606557
52 0.145161290322581
53 0.142857142857143
54 0.140625
55 0.138461538461538
56 0.136363636363636
57 0.134328358208955
58 0.132352941176471
59 0.130434782608696
60 0.128571428571429
61 0.126760563380282
62 0.125
63 0.123287671232877
64 0.121621621621622
65 0.12
66 0.118421052631579
67 0.116883116883117
68 0.124230769230769
69 0.122539307844708
70 0.120893287864926
71 0.119290902375969
72 0.117730439008586
73 0.116210273835699
74 0.114728865735259
75 0.113284751178832
76 0.111876539408867
77 0.110502907971262
78 0.109162598573038
79 0.107854413237859
80 0.106577210734712
81 0.105329903257364
82 0.104111453334288
83 0.102920870950611
84 0.101757210865304
85 0.100619570108338
86 0.0995070856438694
87 0.0984189321867488
88 0.0973543201607502
89 0.0963124937878939
90 0.0952927292991543
91 0.094294333257647
92 0.0933166409861325
93 0.0923590150913423
94 0.0914208440782439
95 0.0905015410479126
96 0.0896005424731846
97 0.0887173070467238
98 0.0878513145965549
99 0.0870020650644959
100 0.0861690775432772
101 0.0853518893684489
102 0.0937045547088593
103 0.0927389918273779
104 0.0917931249535031
105 0.0908663575190789
106 0.08995811680757
107 0.0890678527738336
108 0.0881950369332829
109 0.0873391613157314
110 0.0864997374795652
111 0.0856762955822225
112 0.0848683835032657
113 0.0840755660166045
114 0.0832974240086872
115 0.0825335537397094
116 0.0817835661451025
117 0.0810470861747645
118 0.0803237521676741
119 0.0796132152596973
120 0.0789151388225496
121 0.0782291979320183
122 0.0775550788636783
123 0.0768924786144596
124 0.0762411044485317
125 0.0756006734660754
126 0.0749709121936084
127 0.0743515561946168
128 0.0737423496993287
129 0.0731430452525413
130 0.0725534033784815
131 0.0719731922617478
132 0.0714021874434393
133 0.0708401715316363
134 0.0702869339254461
};
\addlegendentry{sensor 1}
\addplot [semithick, color1]
table {%
0 1
1 0.818181818181818
2 0.75
3 0.692307692307692
4 0.642857142857143
5 0.6
6 0.5625
7 0.529411764705882
8 0.5
9 0.473684210526316
10 0.45
11 0.428571428571429
12 0.409090909090909
13 0.391304347826087
14 0.375
15 0.36
16 0.346153846153846
17 0.333333333333333
18 0.321428571428571
19 0.310344827586207
20 0.3
21 0.290322580645161
22 0.28125
23 0.272727272727273
24 0.264705882352941
25 0.257142857142857
26 0.25
27 0.243243243243243
28 0.236842105263158
29 0.230769230769231
30 0.225
31 0.219512195121951
32 0.214285714285714
33 0.209302325581395
34 0.204545454545455
35 0.2
36 0.195652173913043
37 0.191489361702128
38 0.1875
39 0.183673469387755
40 0.18
41 0.176470588235294
42 0.173076923076923
43 0.169811320754717
44 0.166666666666667
45 0.163636363636364
46 0.160714285714286
47 0.157894736842105
48 0.155172413793103
49 0.152542372881356
50 0.15
51 0.147540983606557
52 0.145161290322581
53 0.142857142857143
54 0.140625
55 0.138461538461538
56 0.136363636363636
57 0.134328358208955
58 0.132352941176471
59 0.130434782608696
60 0.128571428571429
61 0.126760563380282
62 0.125
63 0.123287671232877
64 0.121621621621622
65 0.12
66 0.118421052631579
67 0.116883116883117
68 0.124230769230769
69 0.122539307844708
70 0.120893287864926
71 0.119290902375969
72 0.117730439008586
73 0.116210273835699
74 0.114728865735259
75 0.113284751178832
76 0.111876539408867
77 0.110502907971262
78 0.109162598573038
79 0.107854413237859
80 0.106577210734712
81 0.105329903257364
82 0.104111453334288
83 0.102920870950611
84 0.101757210865304
85 0.100619570108338
86 0.0995070856438694
87 0.0984189321867488
88 0.0973543201607502
89 0.0963124937878939
90 0.0952927292991543
91 0.094294333257647
92 0.0933166409861325
93 0.0923590150913423
94 0.0914208440782439
95 0.0905015410479126
96 0.0896005424731846
97 0.0887173070467238
98 0.0878513145965549
99 0.0870020650644959
100 0.0861690775432772
101 0.0853518893684489
102 0.0937045547088593
103 0.0927389918273779
104 0.0917931249535031
105 0.0908663575190789
106 0.08995811680757
107 0.0890678527738336
108 0.0881950369332829
109 0.0873391613157314
110 0.0864997374795652
111 0.0856762955822225
112 0.0848683835032657
113 0.0840755660166045
114 0.0832974240086872
115 0.0825335537397094
116 0.0817835661451025
117 0.0810470861747645
118 0.0803237521676741
119 0.0796132152596973
120 0.0789151388225496
121 0.0782291979320183
122 0.0775550788636783
123 0.0768924786144596
124 0.0762411044485317
125 0.0756006734660754
126 0.0749709121936084
127 0.0743515561946168
128 0.0737423496993287
129 0.0731430452525413
130 0.0725534033784815
131 0.0719731922617478
132 0.0714021874434393
133 0.0708401715316363
134 0.0702869339254461
};
\addlegendentry{sensor 2}
\end{axis}

\end{tikzpicture}}
	\caption{Results of the experiment with a drift in the ground truth measurement noise for our self-assessment method based on subjective logic.\label{figure:exp02_our_results}}
\end{figure}

As second experiment, we consider a drift in our simulated ground truth measurement noise $\sigma_{w_{gt}}$, which is displayed in Fig.~\ref{figure:exp02_meas_noise}.
Here, the ground truth measurement noise of \textit{sensor~1} drifts from start value of $1$ meter to the end value of $3$ meters.
The underlying process model for the ground truth data generation and for the Kalman filter is again a constant velocity model. 
The results of the time-average NIS including the $95\%$ confidence interval are illustrated in Fig.~\ref{figure:exp02_time_avg_nis_gauss}.
Here, the measure of \textit{sensor~1} gets bigger as the simulated measurement noise gets bigger, while, the time-average NIS of \textit{sensor~2} levels off in the $95\%$ confidence interval. The results of our proposed subjective logic-based method are visualized in Fig.~\ref{figure:exp02_our_results}.
Our self-assessment measure needs approximately the first $40$ time steps in order to clearly separate the performance of the two sensors, but afterwards, the drift is clearly monitored. This effect is reasonable as it can also be seen in Fig.~\ref{figure:exp02_time_avg_nis_gauss} that approximately until time step $40$ both time-average NIS values are within the $95 \%$ confidence interval. 
Our obtained uncertainty of the self-assessment measure is continuously decreasing, which supports the fact that we get more and more certain about our subjective logic-based measure with increasing time.

\subsection{Changes in Process Model}

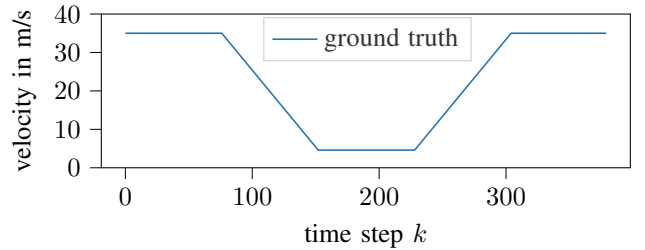
\begin{figure}[b]
	\centering
\begin{tikzpicture}

\definecolor{color0}{rgb}{0.12156862745098,0.466666666666667,0.705882352941177}

\begin{axis}[
height=0.2\textwidth,
width=0.475\textwidth,
legend cell align={left},
legend style={fill opacity=0.8, draw opacity=1, text opacity=1, draw=white!80!black, at={(0.7,0.98)},anchor=north east},
tick align=outside,
tick pos=left,
x grid style={white!69.0196078431373!black},
xlabel={time step $k$},
xmin=-18.95, xmax=397.95,
xtick style={color=black},
y grid style={white!69.0196078431373!black},
ylabel={velocity in m/s},
ymin=0, ymax=40,
ytick style={color=black}
]
\addplot [semithick, color0]
table {%
0 35
1 35
2 35
3 35
4 35
5 35
6 35
7 35
8 35
9 35
10 35
11 35
12 35
13 35
14 35
15 35
16 35
17 35
18 35
19 35
20 35
21 35
22 35
23 35
24 35
25 35
26 35
27 35
28 35
29 35
30 35
31 35
32 35
33 35
34 35
35 35
36 35
37 35
38 35
39 35
40 35
41 35
42 35
43 35
44 35
45 35
46 35
47 35
48 35
49 35
50 35
51 35
52 35
53 35
54 35
55 35
56 35
57 35
58 35
59 35
60 35
61 35
62 35
63 35
64 35
65 35
66 35
67 35
68 35
69 35
70 35
71 35
72 35
73 35
74 35
75 35
76 35
77 34.6
78 34.2
79 33.8
80 33.4
81 33
82 32.6
83 32.2
84 31.8
85 31.4
86 31
87 30.6
88 30.2
89 29.8
90 29.4
91 29
92 28.6
93 28.2
94 27.8
95 27.4
96 27
97 26.6
98 26.2
99 25.8
100 25.4
101 25
102 24.6
103 24.2
104 23.8
105 23.4
106 23
107 22.6
108 22.2
109 21.8
110 21.4
111 21
112 20.6000000000001
113 20.2000000000001
114 19.8000000000001
115 19.4000000000001
116 19.0000000000001
117 18.6000000000001
118 18.2000000000001
119 17.8000000000001
120 17.4000000000001
121 17.0000000000001
122 16.6000000000001
123 16.2000000000001
124 15.8000000000001
125 15.4000000000001
126 15.0000000000001
127 14.6000000000001
128 14.2000000000001
129 13.8000000000001
130 13.4000000000001
131 13.0000000000001
132 12.6000000000001
133 12.2000000000001
134 11.8000000000001
135 11.4000000000001
136 11.0000000000001
137 10.6000000000001
138 10.2000000000001
139 9.80000000000006
140 9.40000000000006
141 9.00000000000006
142 8.60000000000006
143 8.20000000000006
144 7.80000000000006
145 7.40000000000006
146 7.00000000000006
147 6.60000000000006
148 6.20000000000006
149 5.80000000000006
150 5.40000000000006
151 5.00000000000006
152 4.60000000000006
153 4.60000000000006
154 4.60000000000006
155 4.60000000000006
156 4.60000000000006
157 4.60000000000006
158 4.60000000000006
159 4.60000000000006
160 4.60000000000006
161 4.60000000000006
162 4.60000000000006
163 4.60000000000006
164 4.60000000000006
165 4.60000000000006
166 4.60000000000006
167 4.60000000000006
168 4.60000000000006
169 4.60000000000006
170 4.60000000000006
171 4.60000000000006
172 4.60000000000006
173 4.60000000000006
174 4.60000000000006
175 4.60000000000006
176 4.60000000000006
177 4.60000000000006
178 4.60000000000006
179 4.60000000000006
180 4.60000000000006
181 4.60000000000006
182 4.60000000000006
183 4.60000000000006
184 4.60000000000006
185 4.60000000000006
186 4.60000000000006
187 4.60000000000006
188 4.60000000000006
189 4.60000000000006
190 4.60000000000006
191 4.60000000000006
192 4.60000000000006
193 4.60000000000006
194 4.60000000000006
195 4.60000000000006
196 4.60000000000006
197 4.60000000000006
198 4.60000000000006
199 4.60000000000006
200 4.60000000000006
201 4.60000000000006
202 4.60000000000006
203 4.60000000000006
204 4.60000000000006
205 4.60000000000006
206 4.60000000000006
207 4.60000000000006
208 4.60000000000006
209 4.60000000000006
210 4.60000000000006
211 4.60000000000006
212 4.60000000000006
213 4.60000000000006
214 4.60000000000006
215 4.60000000000006
216 4.60000000000006
217 4.60000000000006
218 4.60000000000006
219 4.60000000000006
220 4.60000000000006
221 4.60000000000006
222 4.60000000000006
223 4.60000000000006
224 4.60000000000006
225 4.60000000000006
226 4.60000000000006
227 4.60000000000006
228 4.60000000000006
229 5.00000000000006
230 5.40000000000006
231 5.80000000000006
232 6.20000000000006
233 6.60000000000006
234 7.00000000000006
235 7.40000000000006
236 7.80000000000006
237 8.20000000000006
238 8.60000000000006
239 9.00000000000006
240 9.40000000000006
241 9.80000000000006
242 10.2000000000001
243 10.6000000000001
244 11.0000000000001
245 11.4000000000001
246 11.8000000000001
247 12.2000000000001
248 12.6000000000001
249 13.0000000000001
250 13.4000000000001
251 13.8000000000001
252 14.2000000000001
253 14.6000000000001
254 15.0000000000001
255 15.4000000000001
256 15.8000000000001
257 16.2000000000001
258 16.6000000000001
259 17.0000000000001
260 17.4000000000001
261 17.8000000000001
262 18.2000000000001
263 18.6000000000001
264 19.0000000000001
265 19.4000000000001
266 19.8000000000001
267 20.2000000000001
268 20.6000000000001
269 21
270 21.4
271 21.8
272 22.2
273 22.6
274 23
275 23.4
276 23.8
277 24.2
278 24.6
279 25
280 25.4
281 25.8
282 26.2
283 26.6
284 27
285 27.4
286 27.8
287 28.2
288 28.6
289 29
290 29.4
291 29.8
292 30.2
293 30.6
294 31
295 31.4
296 31.8
297 32.2
298 32.6
299 33
300 33.4
301 33.8
302 34.2
303 34.6
304 35
305 35
306 35
307 35
308 35
309 35
310 35
311 35
312 35
313 35
314 35
315 35
316 35
317 35
318 35
319 35
320 35
321 35
322 35
323 35
324 35
325 35
326 35
327 35
328 35
329 35
330 35
331 35
332 35
333 35
334 35
335 35
336 35
337 35
338 35
339 35
340 35
341 35
342 35
343 35
344 35
345 35
346 35
347 35
348 35
349 35
350 35
351 35
352 35
353 35
354 35
355 35
356 35
357 35
358 35
359 35
360 35
361 35
362 35
363 35
364 35
365 35
366 35
367 35
368 35
369 35
370 35
371 35
372 35
373 35
374 35
375 35
376 35
377 35
378 35
379 35
};
\addlegendentry{ground truth}
\end{axis}

\end{tikzpicture}
	\caption{Changes in the ground truth velocity progression of the data.\label{figure:exp03_gt_velocity}}
\end{figure}

\begin{figure}
	\centering
	\input{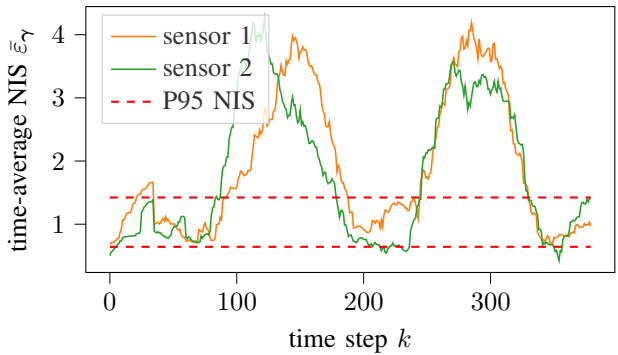}
	\caption{Time-average NIS of the experiment with changes in the ground truth process model.\label{figure:exp03_time_avg_nis_gauss}}
\end{figure}

\begin{figure}
	\centering
	\subfloat[Self-assessment measure.\label{figure:exp03_dc_gauss}]{
	\input{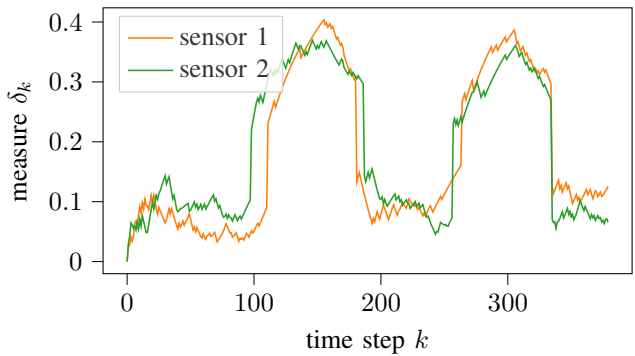}}
  	\\
	\subfloat[Uncertainty of the self-assessment measure.\label{figure:exp03_uncertainty_gauss}]{
		\centering
		\input{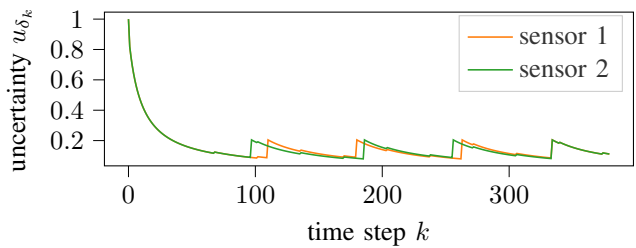}}
	\caption{Results of the experiment with changes in the ground truth process model for our self-assessment method based on subjective logic.\label{figure:exp03_our_result}}
\end{figure}

As last experiment, we simulate changes in the underlying process model of our simulated ground truth data. The changes in our simulated velocity are displayed in Fig.~\ref{figure:exp03_gt_velocity}.
First, we consider a constant velocity of $35$ meters per second, which matches our Kalman filter's assumptions of the process model type. Then, the simulated velocity decreases down to the value of $5$ meters per second which descriptively means that the target brakes. After a section with constant velocity of $5$ meters per second, we accelerate again up to $35$ meters per second.
The calculated time-average NIS of this scenario is depicted in Fig.~\ref{figure:exp03_time_avg_nis_gauss}.
In the sections of constant velocity, the consistency values are mostly within the $95\%$ confidence interval. For the braking and acceleration sections, the consistency values of the two sensors are violated and outside of the confidence interval. Compared to the other experiments, the time-average NIS values are in general smaller, which results from a higher chosen process noise in the Kalman filter's assumptions in order to better visualize the important aspects of this scenario.
The results of our proposed method are shown in Fig.~\ref{figure:exp03_our_result}.
Compared to the time-average NIS, our self-assessment measure is again more consistent when considering the first braking phase. In Fig.~\ref{figure:exp03_time_avg_nis_gauss}, the time-average NIS has two peaks at slightly different locations for \textit{sensor~1} and \textit{sensor~2}. Actually, in our proposed method, this effect is also slightly visible by the peaks in the uncertainty at different time steps, but our self-assessment measure is generally smoother in this braking phase.
Furthermore, our proposed self-assessment measure shows sharper edges when the velocity begins to decrease and increase. Additionally, the peaks in our uncertainty measure in Fig.~\subref{figure:exp03_uncertainty_gauss} supports the recognition of the changes in the velocity progression as explained before.

\section{Conclusion} \label{section:conclusion}

In this contribution, we proposed a self-assessment online method in Kalman filtering based on subjective logic theory. In contrast to classical consistency measures, such as the NIS, we are not only able to obtain a self-assessment online measure of the correctness of the Kalman filter's assumptions, but we are also able to obtain an explicit uncertainty. The latter states how certain we are about the calculated self-assessment measure. As evaluated through simulated data, our proposed method is able to compete with a time-average NIS approach and shows even superior results in some addressed aspects.

In our future work, we aim to implement an adaptive Kalman filter, which is based on our proposed online self-assessment algorithm. Due to the additionally obtained explicit uncertainty and the closed-form algorithm in subjective logic theory, we claim to be able to use subjective logic operators in order to obtain more accurate Kalman filter estimation results. 
Furthermore, we intend to investigate self-assessment of multi-target tracking algorithms using subjective logic.

\bibliographystyle{IEEEtran}
\bibliography{references}

\end{document}